# Ultrafast collapse of molecular polaritons in photoswitch-nanoantennas at room temperature

Joel Kuttruff [a,¶], Marco Romanelli [b,¶], Esteban Pedrueza-Villalmanzo [c,d,¶], Jonas Allerbeck [a,e], Jacopo Fregoni [f], Valeria Saavedra-Becerril [d], Joakim Andréasson [d], Daniele Brida [g], Alexandre Dmitriev [c,*], Stefano Corni [b,h,*], and Nicolò Maccaferri [g,i,*]

[a] Department of Physics, University of Konstanz, 78457, Konstanz, Germany
[b] Department of Chemical Sciences, University of Padova, via Marzolo 1, 35131 Padova, Italy
[c] Department of Physics, University of Gothenburg, Universitetsplatsen 1, 405 30, Gothenburg, Sweden
[d] Department of Chemistry and Chemical Engineering, Chalmers University of Technology, Kemigården 4, 412 96 Göteborg, Sweden
[e] nanotech@surfaces Laboratory, Empa, Swiss Federal Laboratories for Materials Science and Technology, Überlandstrasse 129, 8600 Dübendorf, Switzerland
[f] Department of Physics, Universidad Autónoma de Madrid, Ciudad Universitaria de Cantoblanco, 28049 Madrid, Spain
[g] Department of Physics and Materials Science, University of Luxembourg, 162a avenue de la Faïencerie, L-1511 Luxembourg, Luxembourg
[h] CNR Institute of Nanoscience, via Campi 213/A, 41125 Modena, Italy
[i] Department of Physics, Umeå University, Linnaeus väg 24, 90187 Umeå

¶These authors contributed equally

*alexd@physics.gu.se; *stefano.corni@unipd.it; *nicolo.maccaferri@umu.se

**Molecular polaritons are hybrid light-matter states that emerge when a molecular transition strongly interacts with photons in a resonator. At optical frequencies, this interaction unlocks a way to explore and control new chemical phenomena at the nanoscale. Achieving such a control at ultrafast timescales, however, is an outstanding challenge, as it requires a deep understanding of the dynamics of the collectively coupled molecular excitation and the light modes. Here, we investigate the dynamics of collective polariton states, realized by coupling molecular photoswitches to optically anisotropic plasmonic nanoantennas. Pump-probe experiments reveal an ultrafast collapse of polaritons to a single-molecule transition triggered by femtosecond-pulse excitation at room-temperature. Through a synergistic combination of experiments and quantum mechanical modelling, we show that the response of the system is governed by intramolecular dynamics, occurring one order of magnitude faster with respect to the unperturbed excited molecule relaxation to the ground state.**

## Introduction

Hybrid light-matter polaritonic states arise as a consequence of a coherent energy exchange between the confined electromagnetic field in resonators and the radiating transitions in molecules or, more in general, quantum emitters. The associated strong modification of the energy levels offers tantalizing opportunities of tuning various fundamental properties of matter, such as molecular chemical reactivity or electrical conductivity. As such, the so-called strong coupling regime holds a key potential in a broad range of fields, such as all-optical logic **[1-2],** lasing **[3]**, superfluidity **[4]**, chemistry **[5]** and quantum computing **[6-7]**. The fundamental requirement for strong coupling is in boosting the light-emitter interaction to such an extent that the coherent energy exchange between light and emitters becomes greater than the individual decay rates. Such boost can be achieved by either resorting to a large number of emitters or by confining the electromagnetic field to sub-wavelength volumes. For the former, the most common resonators are photonic cavities, where the electromagnetic field is confined by metallic mirrors (e.g, Fabry-Pérot cavities **[8]** or multilayer heterostructures **[9]**), and polaritons emerge as collective excitations between the light modes and the ensemble of emitters (~$10^6$ to $10^{10}$ molecules). For the latter, sub-wavelength confinement of light has been achieved by exploiting plasmonics **[10]**. Such approach provide near-field enhancement of the electromagnetic field via localized surface plasmon resonances (LSPRs) with an effective mode volume of 1-100 nm$^3$ **[11-12]**, thus allowing the formation of polaritons even with a relatively limited number of emitters (from a single emitter to $10^3$). Due to the broad range of potential applications, the nature of polaritonic states formation and dynamics have been extensively researched over the last decade **[13]**. Coherent time-domain control of the reversible energy exchange between photons and matter, referred to as Rabi oscillations, has been demonstrated in J-aggregates and metal nanostructures **[14],** and for semiconductor quantum wells in a microresonator **[15]**. As well, various incoherent pathways to modify the polaritonic states on ultrafast timescales have been investigated, including charge transfer **[16-17]**, saturation of semiconductor transitions **[18]**, and ground-state bleaching in molecular systems **[19-21]**.

The emerging branch of chemistry using strong coupling to modify the chemical reactions is referred to as polaritonic chemistry **[22]**. There, polaritonic states have been applied to selectively suppress or enhance chemical reactions both in the ground and the excited states **[23-26]**, opening up new chemical reaction pathways, including, among others, remote chemistry **[27]**, singlet fissions **[28]**, and selective isomerization **[29-31]**. Such manipulation of photochemistry makes use of the strong coupling with light to rearrange the electronic energy levels of molecules **[32-33]**. Achieving such control on ultrafast timescales promises many emerging applications combining ultrafast optics and light-driven chemistry **[34]**. This is exceptionally motivating in the context of molecular photoswitches that have already shown potential for ink-less paper **[35]**, stimuli-responsive materials **[36]**, self-healing polymers **[37]** and all-optical switching **[38]**, due to the ability to externally alter their molecular structure by light. One of the most famous photoswitches is spiropyran, which exist in a spiro (SP) and a merocyanine (MC) isomer. UV light triggers the SP→MC isomerization, whereas the reverse reaction is induced by visible light. These compounds display intriguing properties as ultrafast molecular photoswitches, due to the sub-ps kinetics associated with their molecular interconversion **[39]**, and might be key elements to implement future all-optical molecular transistor technologies. The signature feature of the MC form is the emergence of a strong $\pi - \pi^*$ absorption resonance in the visible spectrum **[40]**. While the evidence that strong coupling with the $\pi - \pi^*$ transition modifies the SP-MC photoconversion rate is a milestone for polaritonic chemistry **[32]**, the opportunities for this system at ultrafast timescales are still unexplored.

Here, we devise an archetypical platform capable of selectively accessing the weak and strong coupling regimes and follow the polariton dynamics after impulsive femtosecond-laser excitation. The platform consists in an array of two-mode anisotropic plasmon antennas and spiropyran photoswitches converting to MC form via continuous UV irradiation. To track the role of the coherent plasmon-molecules interaction on ultrafast timescales, we follow the time-evolution of polaritonic states with pump-probe experiments, where the dynamics of the strong coupling is referenced to the weak coupling in the exact same system. Quantum simulations, comprising a novel theoretical framework based on extending the original Tavis-Cummings Hamiltonian **[41],** allow us to interpret the experimental findings assuming changes in the fundamental properties of the coupled system. Our experimental and theoretical analysis reveals an ultrafast modification of the polaritonic state composition, identifying the main relaxation channel as the localization of the initial polaritonic coherent excitation on a molecular excitation. Intramolecular dynamics leads to sub-ps changes of the polaritonic state manifold, one order of magnitude faster than expected from the pure transition time of excited molecules back to the ground state. Revealed sub-ps timescale control of the chemical energy landscape is crucial to advance polaritonic chemistry to the ultrafast regime.

## Results

We employ anisotropic aluminum nanoellipse antennas (see 'Methods' and Supplementary Fig. S1 for fabrication details), displaying two orthogonal spectrally separated LSPRs. A polystyrene nanofilm containing the spiropyran molecules (initially in the SP isomeric form) is spin-coated on top of the nanoantennas and subsequently irradiated with UV light to photo-isomerize the molecules to the MC configuration. The nanoantennas are designed such that the MC molecular absorption is resonant with the long axis LSPR, whereas it is detuned from the short axis LSPR, giving rise to strong and weak coupling regimes, respectively. The steady state optical response of the hybrid system is shown in Fig. 1 for excitation along the long axis (sketched in Fig. 1a) and short axis (sketched in Fig. 1b) of the nanoantennas, respectively. While only the plasmon contribution is visible for the long axis case before UV irradiation (no MC isomers present, orange curve in Fig. 1c), lower (LP) and upper (UP) polaritonic states immediately emerge (black curve in Fig. 1c) upon UV-induced photoconversion of SP to MC, with the corresponding activation of the $\pi - \pi^*$ MC molecular transition. This is a characteristic signature of the system entering the strong coupling regime **[42-43]**. Conversely, when excited along the short axis (Fig. 1d), the spectrum of the coupled plasmon-MC system (black curve) is governed by the superposition of LSPR (pink curve) and the molecular absorption. In this case, the plasmon with low extinction efficiency is out-of-resonance with the molecular transition, leading to two spectral peaks simply superimposing on each other. The absence of notable shifts of the absorption peaks with respect to the bare states in this case signals the weak-coupling regime. To distinguish between the two regimes, we analyze with a full quantum model the modification of the energies of molecular electronic states due to the interaction with the nanoantenna plasmon modes. First, in order to simulate the SP and MC linear absorption spectra (experimental spectra are shown in the inset in Fig. 1d, simulations can be found in Supplementary Fig. S2), the respective molecular ground state geometries are optimized (see Supplementary Fig. S3), and the vertical excitations energies to the first eight excited states of each isomer are calculated. The $S_1$ excited state of MC is then optimized at the same level of theory (in a solvent mimicking the polymer matrix), yielding the relaxed configuration of MC in the $S_1$ state minimum (see Supplementary Fig. S4). The electronic energy associated to this point is about 0.4 eV lower than the vertical excitation energy, meaning that the molecule is not in its equilibrium configuration upon vertical excitation from $S_0$ to $S_1$. Interactions between different MC molecules and the nanoantennas is evaluated by means of an ad-hoc quantum model, which extends the original Tavis-Cummings (TC) Hamiltonian **[41,44]** by explicitly quantizing the modes of arbitrarily shaped nanoparticles **[45]**. The choice of this modelling strategy comes from the fact that the nanoantennas do not display any sophisticated geometrical features that may lead to "picocavity" formation **[46]** and the associated single-molecule strong coupling effects. Hence, investigating collective molecular phenomena while retaining the geometrical shape of the nanoantennas is required in this case. The calculations (Figs. 1e and 1f) corroborate the interpretation of the experimental spectra (more details on the theoretical model can be found in the "Methods" section and in the Supplementary Information).

After assessing that the system can enter the strong coupling regime by driving the long axis LSPR, pump-probe measurements are performed on the hybrid system in both weak and strong coupling regime, and referenced to a UV-exposed film (to trigger the SP→MC isomerization) of pristine spiropyran molecules, i.e., without the presence of the nanoantennas.

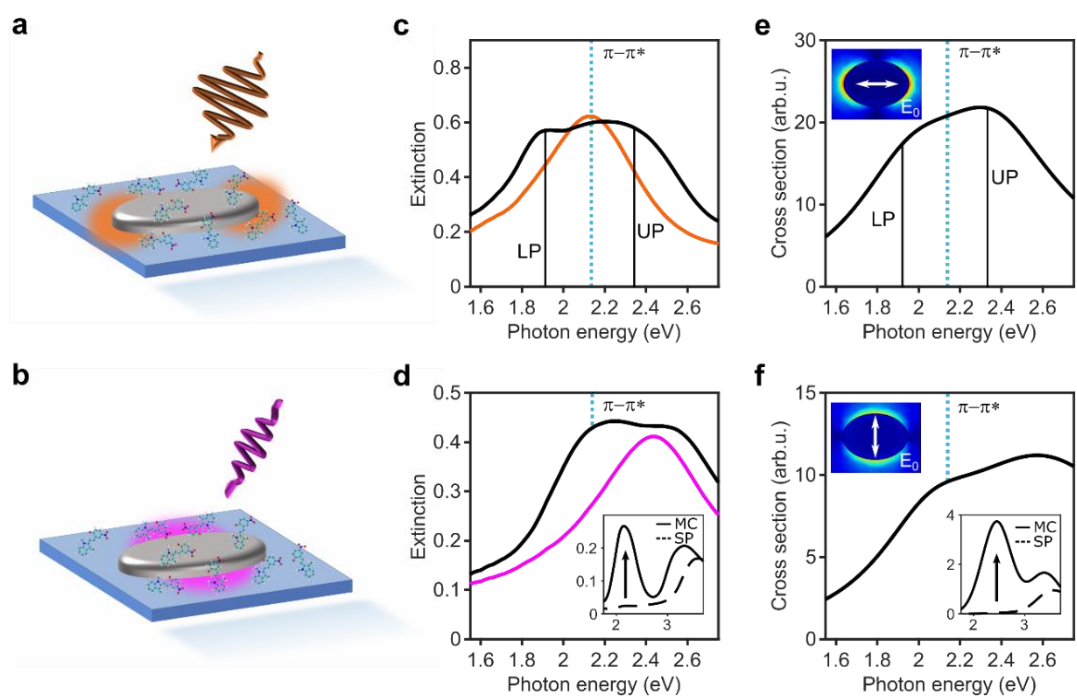

**Figure 1. Hybrid system consisting of photo-switchable molecules and aluminum nanoellipse. a,b** Sketch of the localized plasmon dipolar excitation along the long (a) and short (b) axes of the nanoantenna. The plasmon near-fields are schematically indicated by hot spots (orange and pink) in the sketch. **c,d** Experimental extinction spectra for the long axis (c) and the short axis (d) before (colored) and after (black) UV-induced photoswitching of the spiro isomer (SP) to the merocyanine isomer (MC). The inset in (d) shows the absorption of a molecular film before (dotted) and after (solid) UV irradiation, where the $\pi-\pi^*$ molecular absorption emerges at 2.15 eV (also indicated by the blue dotted lines). **e,f** Extinction cross sections obtained from our model Hamiltonian for long (e) and short (f) axis excitation of the hybrid system. Bottom inset in (f) shows the simulated molecular absorption cross section obtained after ground state optimizations of the molecular structure in the SP (dotted) and MC (solid) configurations. Top insets show confined electric near-fields at both plasmonic resonances calculated using the finite elements method.

The ultrafast response in all the three cases (pristine molecules, molecules weakly coupled and molecules strongly coupled to the nanoantennas LSPRs) is studied using a two-color pump-probe scheme based on a broadly tunable fs-laser spectroscopy platform operating at multi-kHz repetition rate **[47]**. Pump pulses with a temporal duration of 50 fs are tuned to photon energy of 2.3 eV, that is above the $\pi-\pi^*$ transition of the MC molecular form. Time delayed broadband probe pulses span a spectral range from 1.75 eV to 2.5 eV and are compressed to below 20 fs duration. The pump-induced change of probe transmission ΔT/T is monitored for various time delays Δt between pump and probe pulses (see 'Methods' for more details). We first studied the transient transmission of the bare MC film (see Fig. 2). Fig. 2a shows ΔT/T as a function of Δt and the probe photon energy. For positive Δt, a broadband positive signal can be observed, indicating a transient bleaching of the molecular absorption, as previously observed for instance in Ref. **[48]**.

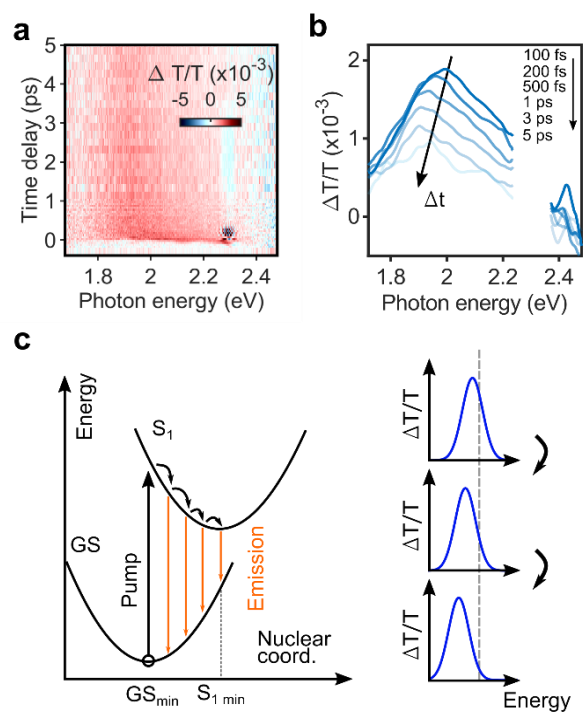

**Figure 2. Ultrafast dynamics of the MC isomer. a** Transient transmission signal as function of time delay between pump and probe pulses and probe photon energy. **b** Pump-probe spectra at increasing pump-probe delay (top to bottom) indicating a redshift of ΔT/T. **c** Scheme of the wave packet relaxation dynamics on the $S_1$ excited state after the pump pulse. The red-shifting positive ΔT/T signal experimentally observed in panel (b) can be interpreted as stimulated emission from the $S_1$ surface while the system is relaxing towards the closest excited state minimum, $S_1$ min (left panel). Resulting ΔT/T spectra are schematically shown in the right panel.

Such bleaching provides a direct measurement of the $S_1$ excited state population at each time, since less photons can be absorbed by the remaining molecules in the MC ground state. In Fig. 2b, we plot spectral cuts of the 2D map shown in Fig. 2a. A clear red-shifting of ΔT/T with increasing Δt can be observed. This can be explained by probe-stimulated photoemission taking place during vibrational relaxation from the Franck-Condon point (where the molecule is vertically excited) toward the $S_1$ minimum, as sketched in Fig. 2c. The left panel in Fig. 2c shows the excitation of MC into the Franck-Condon configuration of the $S_1$ state. The molecular structure then relaxes to a lower energy configuration, subsequently leading to the observed redshift due to the stimulated emission back into the ground state (right panel). From the experiment, we find a maximum spectral shift of 0.3 eV, consistent with the value obtained theoretically (0.4 eV) from full structural relaxation of the $S_1$ state (see also Supplementary Fig. S4).

In Fig. 3, we show the ultrafast dynamics of the hybrid system upon light irradiation polarized either along the long or short axes of the nanoantennas. The transient transmission for the long-axis pumping is shown in Fig. 3a as a function of Δt and the probe photon energy. For positive time delays, we observe a characteristic positive-negative-positive (red-blue-red) spectral shape. Furthermore, our time and energy resolutions allow to track the spectral evolution of ΔT/T, as shown in Fig. 3b, where spectral cuts of the pump-probe 2D map (Fig. 3a) are plotted for increasing values of Δt. As indicated by the black arrow in Fig. 3b, we observe a blue shift of the negative ΔT/T peak within 1 ps after the optical excitation. The observed positive-negative-positive transient spectral lineshape was previously reported, for example, in molecules vibrationally coupled to a cavity **[19]** or in an excitonic transition coupled to a plasmonic lattice **[16]**, and can be interpreted as a shift of the absorption peaks towards each other, and thus a reduction of the spectral separation of the UP and LP bands. While this may be intuitively understood as a pump-induced decrease of the number of molecules effectively coupled to the mode – thus a reduction of the Rabi splitting – explaining the time resolved complex dynamics as shown in Fig. 3b requires a more advanced analysis. After impulsive optical excitation, the upper polariton states of the hybrid system are populated and quickly de-phase within few-tens of femtoseconds, well below the time resolution of our experiment (≈ 50 fs). The excitation then collapses to a single molecule transition, resulting in a system composed by N-1 molecules in their ground state and a single molecule relaxing on its $S_1$ excited state. Being in their ground state, the $S_0 \rightarrow S_1$ electronic transition for the N-1 molecules is still resonant with the long axis plasmon; instead, the relaxation of the individual molecule on its excited state entails a red-shift of the stimulated emission (we note that this is the typical composition that can be obtained for a dark state). The polaritonic energy landscape is thus transiently modified by the intramolecular dynamics of the individual molecule, observable as a transient change of the transmission probed by the delayed optical pulse. To interpret the experimental results, we simulated the transient signal by taking the difference between the absorption spectrum of the 1-photon-space polaritons (shown in Fig. 1), which is the transient spectral signature of ground state bleaching, and the signal coming from the localized red-shifting excited molecular state. The latter term requires computing both the stimulated emission (SE) to the ground state and the absorption toward the 2-photon-space polariton manifold (excited state absorption, ESA) **[49]**. Here, we perform the 2-photon-space calculations (see Methods for additional details) for increasing energy shift of the molecular transition, thus resembling the vibrational relaxation of the molecular structure. Indeed, such theory can reproduce well both the positive-negative-positive motive of the experimental transient spectra and the slight blue shift of the negative peak at increasing Δt, as we show in Fig. 3c. The observed spectral evolution can therefore be assigned to a molecular vibrational relaxation that affects the composition of the polaritonic states manifold.

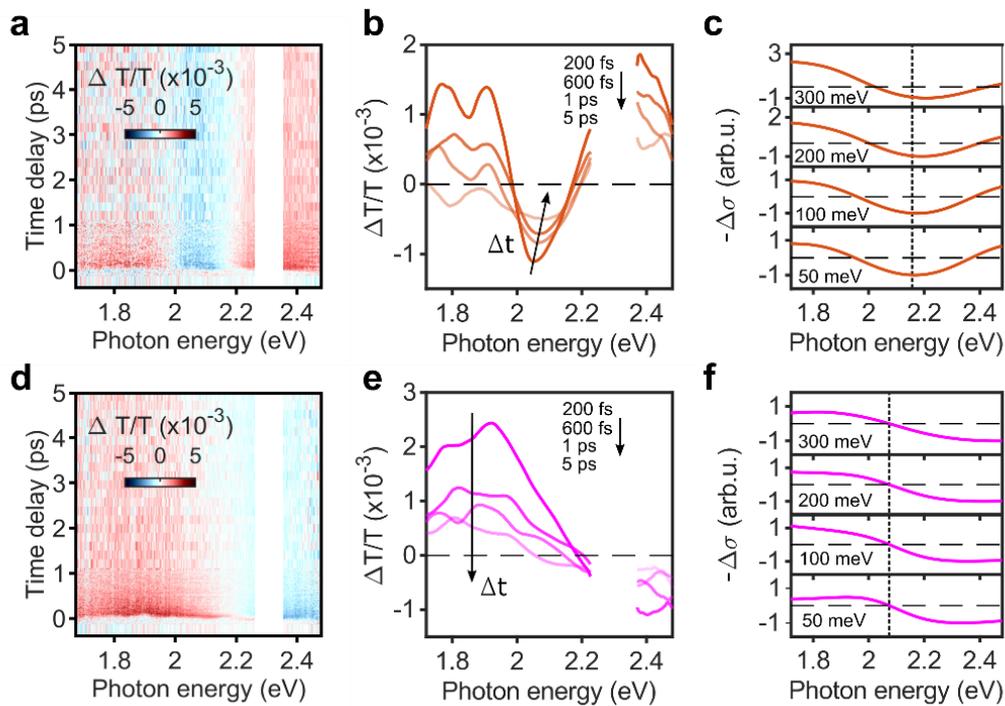

**Figure 3. Ultrafast dynamics of the hybrid system in the strong and weak coupling regimes. a** Transient transmission as function of time delay between pump and probe pulses and probe photon energy when excited along the long axis of the nanoantennas, where the longitudinal plasmon frequency is perfectly tuned to the excitation energy of the MC isomer. **b** Spectral cuts of the pump-probe map in (a) at 200 fs, 600 fs,1 ps and 5 ps. **c** Simulated transient response for the long-axis pumping case. Spectral dynamics are well reproduced by red shifting the molecular transition of one MC (see sketch in Fig. 4c) by 50 meV, 100 meV, 200 meV and 300meV. **d-f** Transient transmission, spectral cuts and simulated transient response for the excitation along the short axis of the nanoantennas, where the MC isomer are off to the plasmon resonance and one of them is gradually red-shifted by 50 meV, 100 meV, 200meV and 300 meV. Here, the excited state relaxation does not influence the dynamics of the weakly coupled system, due to the large initial frequency mismatch with the plasmon resonance.

A simplified sketch describing the resulting rearrangement of electronic energy levels upon optical pumping is depicted in Fig 4. The origin of the positive-negative-positive motive follows from what was already anticipated above: the arrival wavefunction following the probe absorption is, to a first approximation, a product state consisting of the excited molecular state and a 1-photon polariton of the remaining N-1 molecules, whose UP-LP splitting is therefore reduced compared to the ground state absorption (sketched in Fig. 4a and Fig. 4b), thus leading to the characteristic differential positive-negative-positive ΔT/T transient spectral signature (note that SE is providing a minor contribution to the signal).

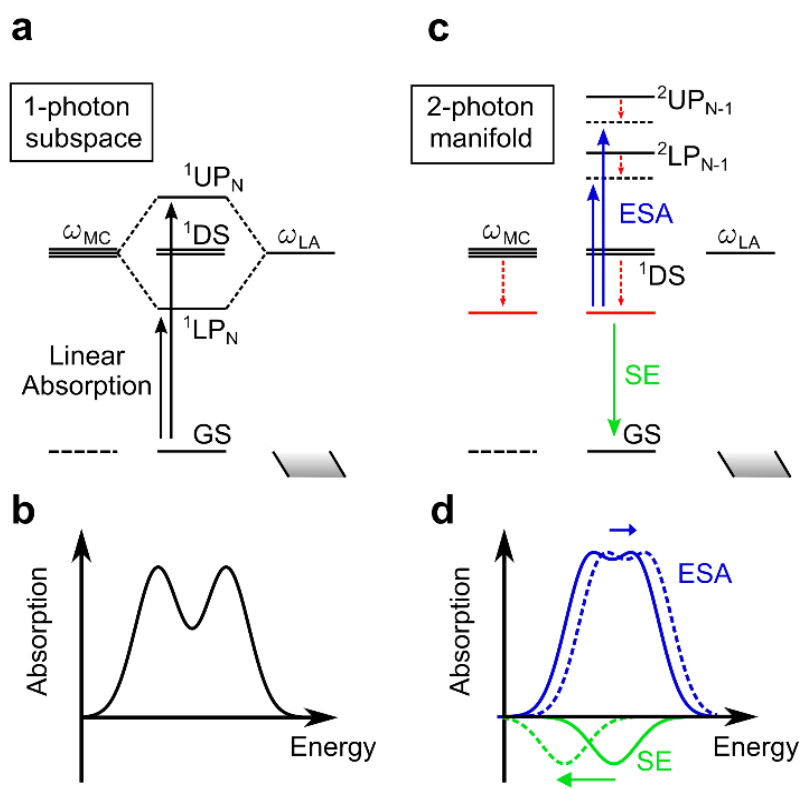

**Figure 4. Visual representation of electronic state reconfiguration after optical pumping in the strong coupling regime. a** Sketch of the electronic energy levels according to the theoretical model before optical pumping of a merocyanince (MC) sample strongly coupled to the long axis plasmon. The 1 superscript in the state labels refer to polaritonic states belonging to the 1-photon subspace obtained through diagonalization of the Hamiltonian in the 1-photon basis. $^1$DS represent molecular dark states. **b** Schematic representation of the linear absorption corresponding to the energy levels shown in panel (a). **c** Sketch of the electronic energy levels after optical pumping. The pump mostly populates $^1$UP at early times, which quickly dephases (≈ tens of fs), collapsing to a single localized MC excited state, which was originally contributing most to $^1$UP state (red energy level). Subsequently, that state undergoes vibrational relaxation (red dashed arrows), during which the probe can promote both stimulated emission (SE) to the ground state and excited state absorption (ESA) to the 2-photons manifold. The two bright states of that manifold accessible from the localized state, here labelled as $^2$UP$_{N-1}$ and $^2$LP$_{N-1}$, can be mostly seen as product states composed of the localized excited state and the 1-photon UP and LP states due to the other N-1 remaining molecules. Indeed, the energy of those states is less affected by the vibrational relaxation (the magnitude of the dashed arrows represents the energy evolution over time). **d** Schematic representation of the ESA and SE contributions (here shown separately for the sake of clarity) originating from the localized state, and their changes over time indicated as dashed curves. In order to obtain the transient signal ΔT/T (Fig. 3c), the sum of the ESA and SE spectra is subtracted from the computed linear absorption.

Upon vibrational relaxation, the state before probe absorption is decreasing its energy faster than what the arrival 2-photon state does, leading to a slight blue shift of the transient signal (see Fig. 3b and Fig. 3c). This is reasonable, since the starting state is essentially a localized excited state of the molecule, while in the arrival state the excitation is partly delocalized on vibrational unrelaxed molecules, with higher energy. On a side note, we observe that the magnitude of the simulated ΔT/T signal of Fig. 3c increases over time, in contrast to what is being observed experimentally. Such discrepancy is expected and is explained by the lack of population decay in the simulations, which physically may be caused by either stimulated emission or non-radiative decay to the ground state.

In Fig. 3d, the transient transmission in the weak coupling regime (i.e., for the short-axis pumping system) is shown. In contrast to the strong coupling case, we now observe a differential (positive-negative) transient spectral lineshape, and contrary to what is observed for the long axis excitation, here there is no significant spectral evolution of the signal over time, as shown in the spectral cuts of Fig. 3e. The same differential signal is again well captured by the theoretical simulations of Fig. 3f. In this case, due to the initial frequency mismatch between the short-axis nanoantenna resonance and the MC transition, the pump quickly drives the system towards a regime where only N-1 remaining MCs are actually able to couple with the short-axis plasmon mode, already at early times, thus reducing the redshift of the molecular absorption caused by the weak coupling with the plasmon compared to ground state absorption, leading then to the positive-negative spectral feature observed.

We now discuss the timescales of the observed ultrafast dynamics in more detail (see Fig. 5). The transient transmission of the bare MC film is shown in Fig. 5a, where the signal at the probe photon energy of the $\pi - \pi^*$ transition (2.1 eV) is shown by the red circles. The bi-exponential behavior follows from an ultrafast (on femtosecond timescale) spectral shift of ΔT/T and a subsequent picosecond decay. More in detail, the induced stimulated emission from the $S_1$ state quickly redshifts due to vibrational relaxation of the molecules from the Franck-Condon configuration towards the $S_1$ minimum, leading to a sub-ps (200 fs) change of ΔT/T at fixed probe photon energy. At lower probe photon energy of 1.85 eV (blue circles), corresponding to the excited state $S_1$ minimum, only a slower (on picosecond timescale) decay of ΔT/T is observable, related to the transition of the molecules back to the ground state, with a time constant of 12 ps. Interestingly, in the weak coupling case (Fig. 5b), a sub-ps decay appears at the relaxed transition energy of 1.85 eV (blue circles in Fig. 5a). This can be inferred to an initial plasmonic component of the decay, which is due to the weak interaction of the molecular transition with the plasmon. After the localization of the wavepacket on the molecular states, the system decays in the same way as the bare molecules. By subtracting the response of the unperturbed molecules, we can extract this ultrafast component induced by the weak coupling of the molecular transition to the plasmonic excitation (blue diamonds in

Fig. 5b), yielding an exponential relaxation with time constant of 400 fs (dashed line in Fig. 5b). In the strong coupling case (Fig. 5c), the same analysis reveals the ultrafast response at the UP (upper panel in Fig. 5c) and LP (lower panel in Fig. 5c) hybrid states, yielding time constants of 350 fs and 360 fs, respectively. Indeed, the polariton lifetimes of the UP and LP states are very similar, indicating the same relaxation mechanism, i.e., non-radiative decay and stimulated photoemission, for both states, as for example observed in Ref. **[18]** for a similar system. Our combined analysis finds a dominant role of intramolecular dynamics on the ultrafast response of the hybrid photoswitch-nanoantennas and establishes sub-ps control of the polaritonic state manifold that is at least one order of magnitude faster than is expected from the transition of excited MC molecules back to the ground state.

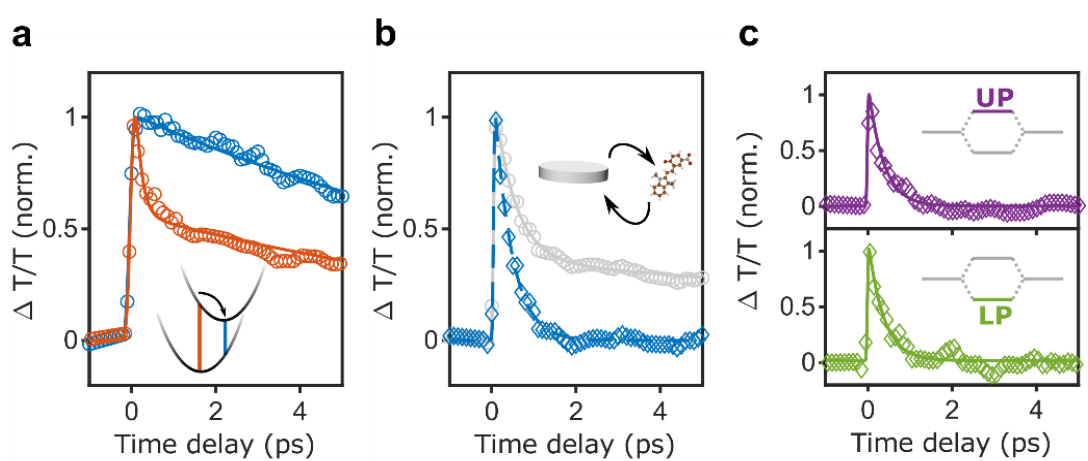

**Figure 5.** Ultrafast decay of the transient signal. **a** Time-resolved cut of the 2D pump-probe data of the bare merocyanine reference film at the energy of the steady-state molecular transition, 2.15 eV (orange), and at the $S_1$ excited state minimum, 1.85 eV (blue). Ultrafast component corresponds to the vibrational relaxation of the molecule, while the slower picosecond component is related to the transition back to the ground state. **b** Cut for the hybrid system in the weak coupling case at the $S_1$ minimum. The grey curve is the raw data and blue curve is the remaining femtosecond decay, obtained by subtracting the slower component shown in (a). **c** Ultrafast decay in the strong coupling case at the energy of the upper and lower polariton states, respectively. The slow component has been subtracted as in (b).

## Conclusion

We presented an in-depth study combining advanced quantum modelling and pump-probe spectroscopy of light-matter coupling in a prototypical photoswitchable molecular-plasmon system stably working at room temperature. The optical anisotropy of the plasmonic resonator allows simultaneous access to weak and strong coupling regimes, depending on the incoming light polarization with respect to the nanoantennas two main axes. In the both the weak and strong coupling regime, experiments show the potential to control the chemical energy landscape of the hybrid system on sub-picosecond timescales, observing a significantly faster dynamics of the excitation with respect to the purely molecular relaxation time. We demonstrate that this effect cannot be explained by a simple plasmon non-radiative decay, which occurs in the first few tens of femtoseconds, but rather originates from the complex intramolecular dynamics within the coupled plasmon-molecules system. A quantum model based on the extension of the Tavis-Cumming Hamiltonian, which can capture the ultrafast dynamics in both the weak and the strong coupling regimes, was developed. This theoretical framework allows also to closely map the ultrafast dynamics to changes of the electronic states in the hybrid system. Our synergistic approach combining ultrafast spectroscopy and advanced quantum modelling paves the way for deep understanding of the ultrafast dynamics in coupled plasmon-molecules systems in general. We believe that our results provide exciting foundation for the further exploration of the synthesis and characterization of strongly coupled photoswitch systems, towards a full control of ultrafast chemical processes at the nanoscale.

**Acknowledgements**

N.M. and D.B. acknowledge support from the Luxembourg National Research Fund (Grant No. C19/MS/13624497 'ULTRON'). N.M., D.B. and S.C. acknowledge support from the European Union under the FETOPEN-01-2018-2019-2020 call (Grant No. 964363 'ProID'). D.B. acknowledges support from the European Research Council through grant no. 819871 (UpTEMPO) and from ERDF Program (Grant No. 2017-03-022-19 'Lux-Ultra-Fast'). J.K. acknowledges the German Research Foundation via SFB 1432. A. D. acknowledges the Swedish Research Council (Grant No. 2017-04828). N.M. acknowledges support from the Swedish Research Council (Grant No. 2021-05784).

# METHODS

<u>Fabrication and steady-state optical characterization.</u> Hole-colloidal lithography process details are described elsewhere **[50]**. Briefly, a sacrificial layer of PMMA is spin-coated on a quartz substrate, with a thickness around 250 nm. After depositing polystyrene (PS) beads of 100 nm-diameter, a Cr mask is evaporated on top while the sample is tilted 45 deg respect to the surface, allowing to cast a shadow on the PMMA surface with an elliptic shape. After tape-stripping the beads from the surface, $O_2$ plasma etching generates holes that allows the deposition of Al by e-beam evaporation. The final metasurface sample is obtained by lift-off the PMMA with acetone in an ultrasonic bath.

SP molecules were mixed with PS polymer in toluene and spin-coated on top of the Al metasurface and leave to evaporate at room temperature, forming a thin film covering the Al ellipses. More details regarding the fabrication are reported in the Supplementary Information. Steady-state optical characterization shown in Figure 1 c,d was performed with white light polarized along the LA and SA of the Al nanoellipses, respectively. To convert the SP to its MC configuration, the sample was continuously illuminated with UV light (365 nm). The photoconversion was monitored by measuring the absorption at the $\pi - \pi^*$ transition (2.1 eV).

<u>Pump-probe spectroscopy.</u> The experimental platform used in this study is based on a commercial Yb:KGW regenerative amplifier system working at a laser repetition rate of 50 kHz was previously described Ref. **[47]**. Pump pulses are generated by a noncollinear optical parametric amplifier (NOPA) working in the visible spectral range, where a bandpass filter is used before the parametric amplification to restrict the spectrum to a central photon energy of 2.3 eV and spectral bandwidth of 50 meV. The pump-induced change of transmission is probed by a white-light supercontinuum generated from sapphire between 1.75 eV and 2.5 eV, which is temporally compressed using dielectric chirped mirrors. The pump pulse energy is set to 20 µJ/cm$^2$ and the probe pulse energy is adjusted to ensure at least a 1:10 energy ratio compared to the pump. Using a spherical mirror with 300 mm focal length, spot sizes of 120 µm and 150 µm are set for probe and pump pulses, respectively. Pump and probe pulses have parallel polarization and interact with the sample at a small noncollinearity angle, allowing to spatially block the pump pulse after sample interaction. Remaining scattering from the pump is then removed from the data in post-processing. A fast spectrometer camera spectrally resolves the probe pulse after sample interaction. For all measurements, the dye molecules are prepared in their MC configuration, as described above. Photoisomerization of some MC molecules during the pump-probe measurements is taken into account.

Numerical calculations. Quantum mechanical calculations of isolated bare SP and MC molecules were performed by means of Gaussian 16 **[51]**, the solvent was considered as an implicit medium through the default Gaussian implementation of the IEF-PCM formalism **[52]**. The choice of using ethylbenzene as implicit solvent to reproduce the polystyrene environment was done because of its close structural resemblance to the styrene molecule (they also present almost identical dielectric constants) and its prompt availability in the Gaussian package. The aluminum nanoellipse was created through the Gmsh code **[53]**, (see Supplementary Fig. S5) and the coupling numerical values were computed through our homemade code TDPlas, implementing the theory described previously **[45]**. More precisely, one single nanoparticle (considering the elliptical shape, see Supplementary Fig. S5) is surrounded by many MC molecules following an elliptical grid pattern (shown in Supplementary Fig. S6), where each molecule is described as a point dipole oriented perpendicularly to the metal surface. Unlike the standard TC Hamiltonian, here we do not assume a constant coupling. Instead, we evaluate the coupling of each molecule with each plasmon mode, depending on its position and orientation w.r.t. the electromagnetic field associated to the plasmonic modes. In addition, we consider simultaneously the two (LA and SA) plasmon modes to include the two individual field distributions (electric near-field calculated via finite elements method shown inset of Figs. 1 e,f). The coupling values for each molecule are numerically evaluated considering a Drude-Lorentz quantized description of the metallic response **[45]**. The eigenvalues and eigenvectors of the resulting Hamiltonian, that is, the energies and state composition of the polaritonic states, are then used to obtain the steady-state spectra by making use of a linear response expression of the polarizability, evaluated as a sum over the polaritonic energy levels. The same Hamiltonian written in the 2-photons state basis (see SI for more details) is then diagonalized to obtain information about the 2-photons-space polaritons and simulate the transient signal observed in the pump-probe experiments shown in Fig. 3. The simulated raw data for the LA-SA case of Fig. 3c-f featured constant deviations from the zero baseline, so a-posteriori constant shifts have been applied to each transient spectrum to obtain the data reported in Fig. 3c-f. More precisely, each simulated curve has been overall positively shifted to make its inflection point aligned with the zero baseline and then has been normalized to the most negative value in that frequency range. The decay rates associated with each polaritonic state are obtained by resorting to a non-Hermitian formulation of the Hamiltonian where energies of the uncoupled states bring an imaginary component which corresponds to either a molecular or plasmonic decay rate (see SI for the values considered) **[54-55]**. The diagonalization of such Hamiltonian directly returns the decay rates of the polaritons as imaginary component of their associated eigenvalues **[56]**. Explicit theory formulation can be found in the Supplementary Information.

# SUPPORTING INFORMATION

## 1. Sample fabrication

### 1.1. Fabrication procedures

Quartz substrates (2 x 2 cm$^2$) were cleaning immersing them consecutively in acetone and isopropanol, in an ultrasonic bath, to be finally rinsed in DI water and dry with nitrogen. Hole-mask colloidal lithography (HCL) **[1]** nanofabrication technique was employed to deposit Al nano-ellipses. Briefly, a PMMA layer (950 PMMA A4) was spin-coated for 60 seconds at 3000 rpm. After a brief O2 plasma treatment (5 s, 250 mTorr at 50 W), an aqueous polyelectrolyte layer of PDDA solution at 0.2% was pipetted into the surface and rinsed with DI water after 1 minute. Then, an aqueous solution of polystyrene (PS) beads of 100 nm-diameter (0.2 % in volume) were pipetted and rinsed with DI water after 2 minutes. A 10 nm Cr layer is then evaporated on top of the tilted substrate, so the shadow pf the PS beads generates an elliptical-hole mask. After the tape-stripping of the PS beads, O2 plasma etching (5 min, 250 mTorr at 50 W) will generate the holes in the PMMA, and 60 nm e-beam evaporated Al will be deposited. Finally, hot acetone in an ultrasonic bath is used in the final lift-off step. The final nanoellipses have a long axis of around 140 nm. The spiropyran (1,3,3-Trimethylindolino-6´-nitrobenzopyrylospiran, from Tokyo Chemical Industry) is mixed with poly(styrene) standard (MW= 45730, Aldrich Chem. Co.) (PS) in a ratio spiropyran:PS of 2:0.5 wt% and dissolved in toluene (Merck). The as-prepared solution is spinned into the Al nanoellipse sample at 3000 rpm for 60 seconds. This generates a transparent thin film with a thickness of around 50 nm, determined by perfilometry (Surface profiler Tencor AS500). The samples were characterized using scanning electron microscopy (SEM) (Zeiss Supra 55VP at 30 kV, see Figure S1) and in tapping mode AFM (Bruker Dimension 3100 SPM).

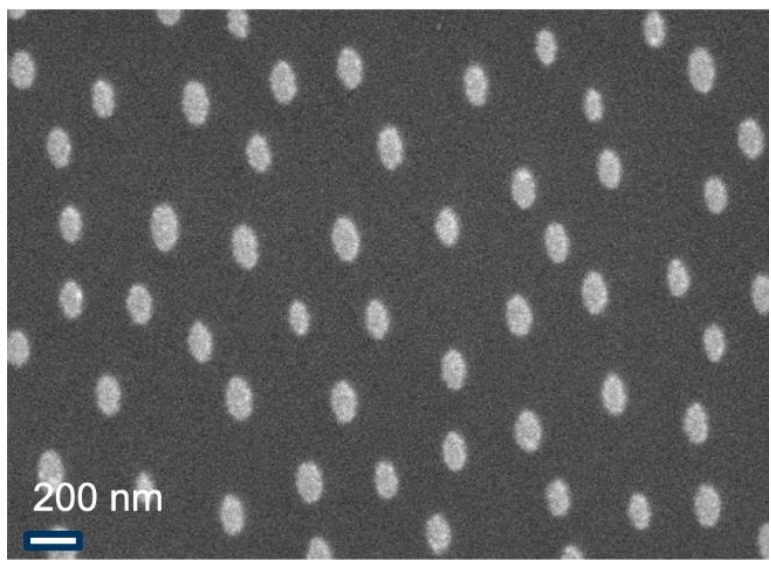

***Figure S1:*** *SEM image of the aluminum nanoellipses prepared by hole-mask colloidal lithography.*

## 2. Computational part

### 2.1. Quantum mechanical calculations of the isolated spiropyran and merocyanine molecules:

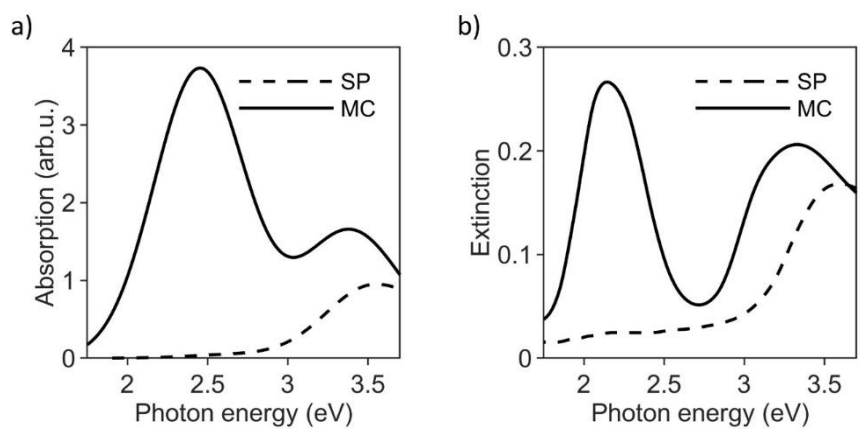

*Figure S2:* a) Simulated linear absorption in Ethylbenzene at TD-DFT/B3LYP 6-31g(d,p). b) Experimental extinction spectra of the polystyrene film containing the molecules.

The ground state structures of the two molecules were optimized at B3LYP/6-31g(d,p) level of theory using the Gaussian 16 package **[2]** in vacuum:

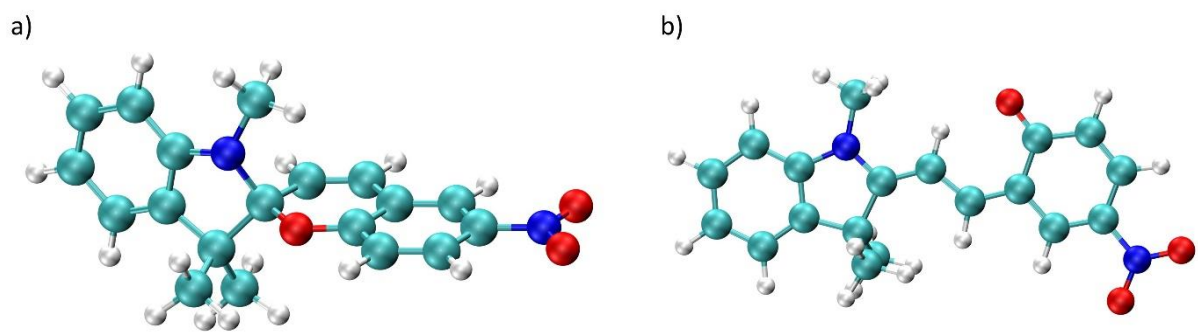

*Figure S3*: a) Ground state optimized structure of the SP isomer. b) Ground state optimized structure of the MC isomer.

From the two structures reported in Figure S3, vertical excitation energies at TD-DFT/B3LYP 6-31g(d,p) for the first 8 excited states were computed, including ethylbenzene as an implicit solvent using the standard IEF-PCM **[3]** implementation of the Gaussian 16 package. The resulting simulated linear absorption (obtained with a Half-Width-Half-Height broadening of 0.333 eV for each transition) is reported above in Figure S2, displaying a very good agreement with the experimental counterpart.

In addition to the ground-state optimized structures, the $S_1$ state of the merocyanine isomer was optimized in solvent to find the closest excited-state minimum towards which the system can relax upon excitation. Following TDDFT/B3LYP 6-31g(d,p), the relaxed $S_1$ structure shown in Figure S4 panel b) is ≈ 0.4 eV more stable than the $S_1$ state at the ground state geometry, thus suggesting that the red-shift of the ΔT/T pump-probe signal reported in the main text Figure 2 (≈ 0.3 eV), can indeed be assigned to stimulated emission of excited MCs relaxing towards the $S_1$ minimum.

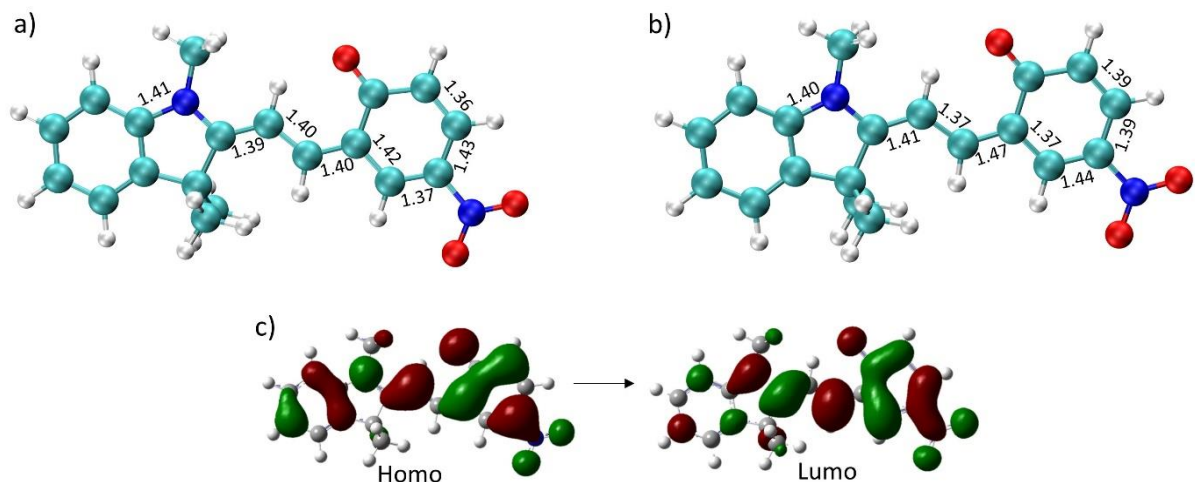

**Figure S4:** *a) Ground state optimized structure of MC (the reported bond lengths are in Å). b) Excited state optimized structure of MC (the reported bond lengths are in Å). c) The $S_0 \rightarrow S_1$ bright transition of the MC (peaking at 2.45 eV in our simulations, see Figure S2) mostly involves the HOMO → LUMO orbitals here reported.*

2.2. **Theoretical model to simulate the optical response of the coupled system**

The nanoellipse employed in the simulations throughout this work was created using the Gmsh code **[4]** (Figure S5). Its associated extinction spectra obtained with different polarizations of the incoming electric field are reported in Figure S8 panel a).

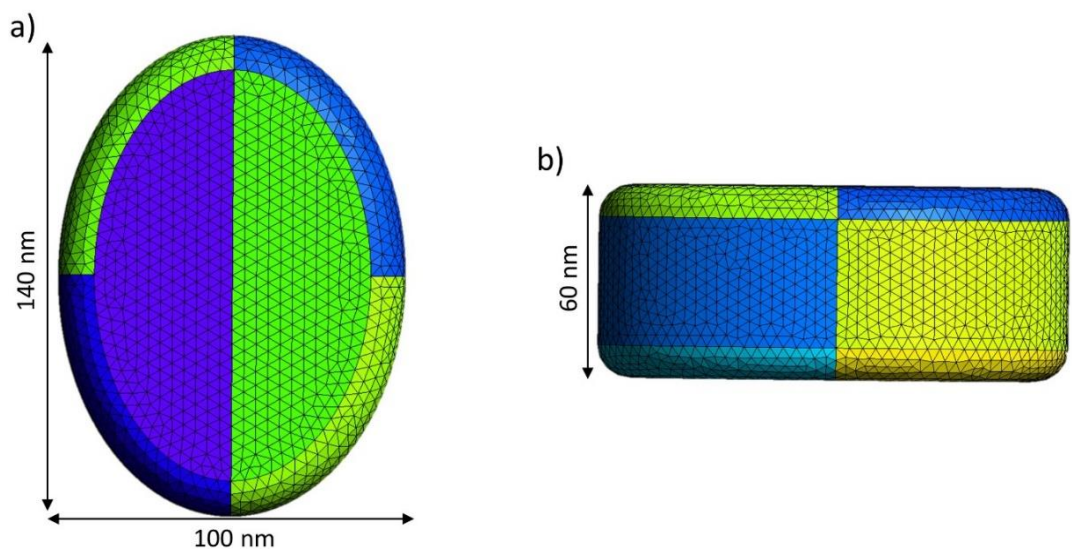

**Figure S5:** *Nanoellipse created with the gmsh code and employed in the simulations: front view a), side view b). The surface elements, called tessera, that are here observable result from the surface mesh discretization operated by the gmsh code and it is necessary to numerically evaluate the coupling parameters with the MC molecules, see details below.*

In order to analyze the optical properties of the hybrid light-matter states (i.e. Polaritons) mentioned in the main text, we employed the quantization scheme of the metallic response based on a single Drude-Lorentz oscillator model of the dielectric function as developed previously **[5]**. In short, starting from a classical description of the nanoparticle within the PCM-NP framework **[6]** subject to an external frequency-dependent perturbation (in the quasi-static limit), one can obtain the following expression:

$$\boldsymbol{q}(\omega) = \boldsymbol{Q}^{IEF}(\omega)\boldsymbol{V}(\omega)$$

*(1)*

Where $\boldsymbol{q}(\omega)$ is a collection of polarization surface charges that account for the nanoparticle linear response under the external electrostatic potential $\boldsymbol{V}(\omega)$ and $\boldsymbol{Q}^{IEF}(\omega)$ is the frequency-dependent response function, that can be recast into a diagonal form as **[7]**:

$$\boldsymbol{Q}^{IEF}(\omega) = -\boldsymbol{S}^{-\frac{1}{2}}\boldsymbol{T}\boldsymbol{K}(\omega)\boldsymbol{T}^{\dagger}\boldsymbol{S}^{-\frac{1}{2}}$$

*(2)*

with

$$K_p(\omega) = \frac{2\pi + \Lambda_p}{2\pi\frac{\varepsilon(\omega)+1}{\varepsilon(\omega)-1} + \Lambda_p}$$

*(3)*

where $\boldsymbol{K}(\omega)$ is the diagonal response matrix derived from eigenvalues $\Lambda_p$ of the appropriate IEF matrix **[7]**, $\varepsilon(\omega)$ is the frequency dependent dielectric function and $\boldsymbol{S}$ is the matrix storing the electrostatic potentials between discrete points of the dielectric.

Based on this classical description and assuming a single Drude-Lorentz oscillator model for the metal dielectric function one can retrieve the following quantum response function**[8]**:

$$Q_{kj}^{quant}(\omega) = -\sum_p \frac{\langle 0|\hat{q}_k|p\rangle\langle p|\hat{q}_j|0\rangle}{\omega_p + \omega + i\Gamma/2} + \frac{\langle p|\hat{q}_k|0\rangle\langle 0|\hat{q}_j|p\rangle}{\omega_p - \omega - i\Gamma/2}$$

*(4)*

where each matrix element k, j of the matrix response function $Q^{quant}(\omega)$ is evaluated on the representative points of the tesserae k and j, and $\omega_p$ is the frequency of a given plasmon mode computed as:

$$\omega_p^2 = \omega_0^2 + \left(1 + \frac{\Lambda_p}{2\pi}\right)\frac{\Omega_p^2}{2}$$

*(5)*

with $\Omega_p$ being the plasma frequency of the bulk metal, and $\Gamma$ is the damping rate of the DL oscillator. It is important to notice that elements as $\langle 0|\hat{q}_k|p\rangle$ represent the quantized transition charges sitting on the k-th tessera related to the mode *p*, so for a given plasmon mode *p* a collection of quantized charges (one for each tessera) is given. These quantities, that are analogous to molecular transition densities, can be used to numerically evaluate the coupling strength between a selected plasmon mode and a given molecule, that reads:

$$\hat{g}_p = \sum_j q_{p,j} \hat{V}_j$$

(6)

Where $\hat{V}_j$ is the molecular electrostatic potential operator evaluated at the position of the j-th tessera where the $q_{p,j}$ charge related to the *p* mode lies on.

Considering the previously mentioned theoretical background, in order to shed light on the optical response of the coupled system the aluminum nanoellipse of Figure S4 was surrounded by many merocyanine molecules (MCs), each one described as a point-dipole and an ad-hoc Tavis-Cummings-like Hamiltonian (in a.u.) was set up:

$$\hat{H} = \sum_{i=1}^{N} \bar{\omega}_{MC} \hat{\sigma}_i^\dagger \hat{\sigma}_i + \bar{\omega}_{LA} \hat{a}_{LA}^\dagger \hat{a}_{LA} + \bar{\omega}_{SA} \hat{a}_{SA}^\dagger \hat{a}_{SA} + \sum_{i=1}^{N} g_{i,LA} (\hat{a}_{LA}^\dagger \hat{\sigma}_i + \hat{a}_{LA} \hat{\sigma}_i^\dagger)$$
$$+ \sum_{i=1}^{N} g_{i,SA} (\hat{a}_{SA}^\dagger \hat{\sigma}_i + \hat{a}_{SA} \hat{\sigma}_i^\dagger)$$

(7)

Where the $\hat{\sigma}_i$ and $\hat{\sigma}_i^\dagger$ are the typical molecular transfer operators and $\hat{a}_{LA}^\dagger$ ($\hat{a}_{LA}$) is the creation (annihilation) operator related to the long-axis mode of the nanoellipse, whereas $\hat{a}_{SA}^\dagger$ ($\hat{a}_{SA}$) is the corresponding operator for the short axis mode. We also note that $\bar{\omega}_{MC} = \omega_{MC} - i\Gamma_{MC}$, $\bar{\omega}_{LA} = \omega_{LA} - i\Gamma_{LA}$ and $\bar{\omega}_{SA} = \omega_{SA} - i\Gamma_{SA}$ where each imaginary component contains the corresponding decay rate of the uncoupled state. The inclusion of the decay rate as an imaginary part results in a non-Hermitian Hamiltonian, previously applied to the case of individual molecules **[9-10]** and more recently to plexcitonic systems **[11]**. The molecular decay rate is set to $\Gamma_{MC} = 1.5 \times 10^{-4}$ $au$ (atomic units) based on the linewidth of the experimental data shown in Figure S2, whereas $\Gamma_{LA} = 0.038$ $au$ and $\Gamma_{SA} = 0.053$ $au$ were chosen on the basis of the fitted DL parameters as described in Figure S8.

In the 1-photon-space basis the Hamiltonian shown above (7) reads:

$$\hat{H} = \sum_{i=1}^{N} \bar{\omega}_{MC} |G_1..E_i..G_N; 0,0\rangle\langle 0,0; G_1..E_i..G_N| + \bar{\omega}_{LA} |G_1..G_N; 1,0\rangle\langle 1,0; G_1..G_N|$$
$$+ \bar{\omega}_{SA} |G_1..G_N; 0,1\rangle\langle 0,1; G_1..G_N| + \sum_{i=1}^{N} g_{i,LA} |G_1..G_N; 1,0\rangle\langle 0,0; G_1..E_i..G_N|$$
$$+ adj. + \sum_{i=1}^{N} g_{i,SA} |G_1..G_N; 0,1\rangle\langle 0,0; G_1..E_i..G_N| + adj.$$

(8)

Where $G_i$ stands for the ground state of the i-th molecule, $E_i$ means the excited state of the i-th molecule and $g_{i,LA}$ ($g_{i,SA}$) is the coupling strength between the i-th molecule $G_i$->$E_i$ transition and the long-axis (short-axis) plasmon mode. Note that unlike the original TC Hamiltonian, here each molecule features its own coupling strength and orientation with respect to the given plasmon modes (two simultaneously considered in this work). The values of $g_{i,LA}$ ($g_{i,SA}$) were calculated for each molecule by the interaction of the $G_i$->$E_i$ transition dipole with the relevant plasmonic mode charges [5].

Diagonalization of such Hamiltonian (8) gives access to the 1-photon-space polaritonic energies (eigenvalues) and corresponding polaritonic wavefunctions (eigenvectors). Transition dipoles from the GS to each 1-photon-space polariton (1PL), that is $\langle GS|\hat{\mu}|1PL\rangle = \vec{\mu}_{1PL}$, were calculated by linear combination of molecular terms obtained by the TDDFT calculations of the previous section, such as $\langle GS|\hat{\mu}|E_1..G_N;0,0\rangle = \vec{\mu_1}$, and plasmonic terms, like $\langle GS|\hat{\mu}|G_1..G_N;1,0\rangle = \vec{\mu}_{LA}$ and $\langle GS|\hat{\mu}|G_1..G_N;0,1\rangle = \vec{\mu}_{SA}$. Such plasmonic transition dipole is evaluated as $\sum_j q_{p,j}\vec{r}_j$, where $\vec{r}_j$ is the position of the center of each tessera where the quantized surface charge $q_{p,j}$ lies on, for a given mode $p$.

The decay rates $\Gamma_{1PL}$ associated to each polaritonic state are directly obtained from the imaginary component of the eigenvalues of the Hamiltonian (8).

Once these quantities are computed, the linear response expression of the polarizability of the entire system (molecule+nanostructure) as a sum over the polaritonic states reads:

$$\alpha_{ij}(\omega) = \sum_{1PL} \frac{\langle GS|\hat{\mu}_i|1PL\rangle\langle 1PL|\hat{\mu}_j|GS\rangle}{\omega_{1PL} + \omega + i\Gamma_{1PL}/2} + \frac{\langle GS|\hat{\mu}_i|1PL\rangle\langle 1PL|\hat{\mu}_j|GS\rangle}{\omega_{1PL} - \omega - i\Gamma_{1PL}/2}$$

*(9)*

with $\langle GS|\hat{\mu}_i|1PL\rangle = \vec{\mu}_{1PL,i}$ being the i-th component of the transition dipole to a given polaritonic state 1PL with energy $\omega_{1PL}$ and damping rate $\Gamma_{1PL}$. Such quantity (9) is used to evaluate the absorption cross section (reported in panels e-f, Figure 1 main text) through [12]:

$$\sigma_{ii}(\omega) = \frac{4\pi\omega}{c} Im\{\alpha_{ii}(\omega)\}$$

*(10)*

where ii is either xx or yy in our simulations for the LA or SA case, respectively. This is because the experiments are done with polarized light pulses, so only some components of the polarizability tensor become relevant.

In order to simulate the pump-probe transient response shown in Figure 3 main text, the 2-photons-space states (2PL) are also required, thus the following Hamiltonian has to be diagonalized:

$$\hat{H} = \sum_{i=1}^{N}\sum_{j>i} 2\bar{\omega}_{MC}|E_iE_j..G_N;0,0\rangle\langle 0,0;E_iE_j..G_N| + \sum_{i=1}^{N}(\bar{\omega}_{LA} + \bar{\omega}_{MC})|E_i..G_N;1,0\rangle\langle 1,0;E_i..G_N|$$

$$+ 2\bar{\omega}_{LA}|G_1..G_N;2,0\rangle\langle 2,0;G_1..G_N| + \sum_{i=1}^{N}\sum_{j>i} g_{j,LA}|E_iE_j..G_N;0,0\rangle\langle 1,0;E_iG_j..G_N|$$

$$+ adj. + \sum_{i=1}^{N} \sqrt{2}g_{i,LA}|E_i..G_N;1,0\rangle\langle 2,0;G_i..G_N|$$

$$+ adj. + corresponding\ terms\ of\ SA\ mode$$

*(11)*

We note that considering the 2-photons-space polaritons, in addition to the more canonical 1-photon-space states, was recently suggested to be a theoretically robust way for interpreting transient pump-probe data of molecules in QED cavities by A. DelPo et al. **[13]**. We also remark that eigenenergies and eigenstates of our Hamiltonian (11) exactly match the analytical results reported in their work in the limit of identical molecules and couplings with all transition dipoles oriented along the direction of the cavity mode.

As mentioned in the main text, the pump-induced excitation of the polaritonic state (see Figures 3-4, main text) leads to the population of one localized molecular state $|MC^*\rangle$ from which stimulated emission (SE) to the GS and excited state absorption (ESA) to the 2-photons-space polaritons transiently take place. The SE spectrum is simulated using (9) and (10), based on the wavefunction of the emitting state $|MC^*\rangle$ obtained through diagonalization of the 1-photon Hamiltonian (8) with one red-shifted MC. Instead, the ESA term can be obtained through equation (10), having:

$$\alpha_{ij}(\omega) = \sum_{2PL} \frac{\langle MC^*|\hat{\mu}_i|2PL\rangle\langle 2PL|\hat{\mu}_j|MC^*\rangle}{(\omega_{2PL} - \omega_{MC^*}) + \omega + i|\Gamma_{2PL} - \Gamma_{MC^*}|/2} + \frac{\langle MC^*|\hat{\mu}_i|2PL\rangle\langle 2PL|\hat{\mu}_j|MC^*\rangle}{(\omega_{2PL} - \omega_{MC^*}) - \omega - i|\Gamma_{2PL} - \Gamma_{MC^*}|/2}$$

(12)

Where the 2-photons-space polaritons are in this case obtained through diagonalization of (11) with all the diagonal elements corresponding to the red-shifted MC molecule modified accordingly to account for the frequency shift.

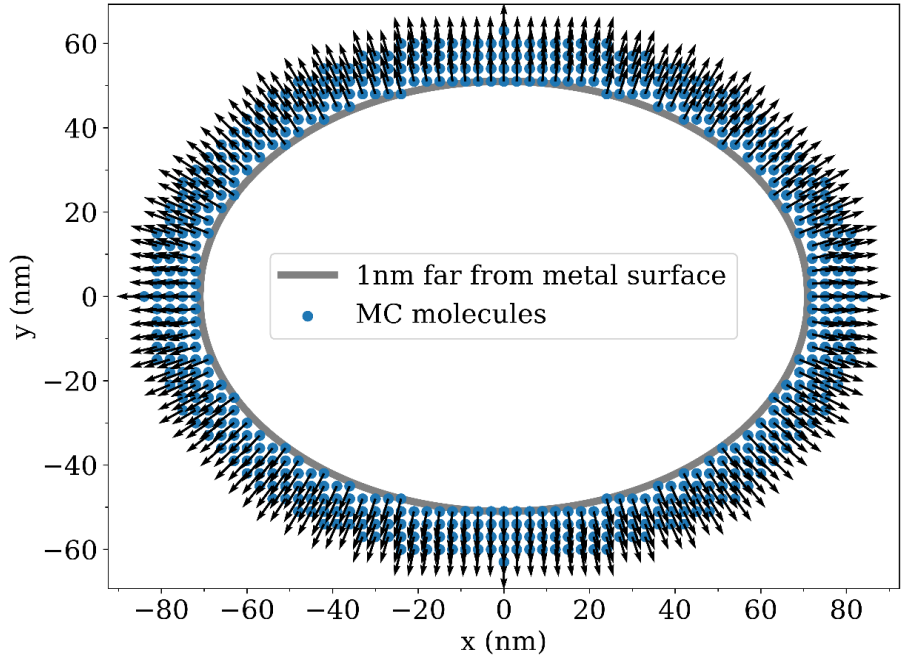

*Figure S6:* *2D slice of the elliptical grid employed in the simulations. Each blue dot represents a MC molecule described as a point-dipole oriented perpendicular to the metal surface. The nearest neighbor distance was set to 3 nm, which corresponds to the sample concentration of 2% wt (≈ 1 molecule per 26 nm³). Data provided in the main text was obtained with such setup on a 2D layer of molecules only, scaling the coupling values accordingly to match with the experimental linear absorption of the coupled system. More detailed information regarding the computed coupling strength with the full 3D grid are reported in Figure S7.*

As described in Figure S6, the data reported in the main text was obtained considering a single 2D grid layer of molecules located at half height of the nanoellipse to save computational time, and the corresponding calculated coupling values have been increased until a good matching with the experimental linear absorption was found (see Figure 1 main text). (We recall the relation between the Rabi splitting energy $\Omega_R$ and the number of molecules $N_{res}$ resonantly coupled to the mode in the simplified case of equal coupling for all the molecules, i.e., $\Omega_R \propto \sqrt{N_{res}}$ ).

Despite this computationally convenient choice, calculations with full 3D grid of MCs surrounding the nanoellipse were also performed in order to assess whether the computed couplings of the real 3D system were large enough to correctly reproduce the experimental data (without any ad hoc increase). As shown in Figure S7 panel a), two distinct grid steps have been tested, 3.0nm (corresponding to 1 molecule per 27 nm$^3$) and 2.0 nm (corresponding to 1 molecule per 8 nm$^3$) respectively, and it turned out that in the tightest 3D grid case (2 nm step size, green curve), by employing a multiplying factor of 2 for the computed couplings, the corresponding Rabi splitting is slightly larger than the experimental one, which perfectly matches with that predicted assuming 1 single 2D layer with step size of 3nm and couplings scaled by a factor of 15 (blue curve). These numerical tests point out that in the realistic limit of a full 3D grid having an intermediate step size between 2 and 3 nm the calculated couplings would reproduce the experimental data, thus corroborating the theoretical model.

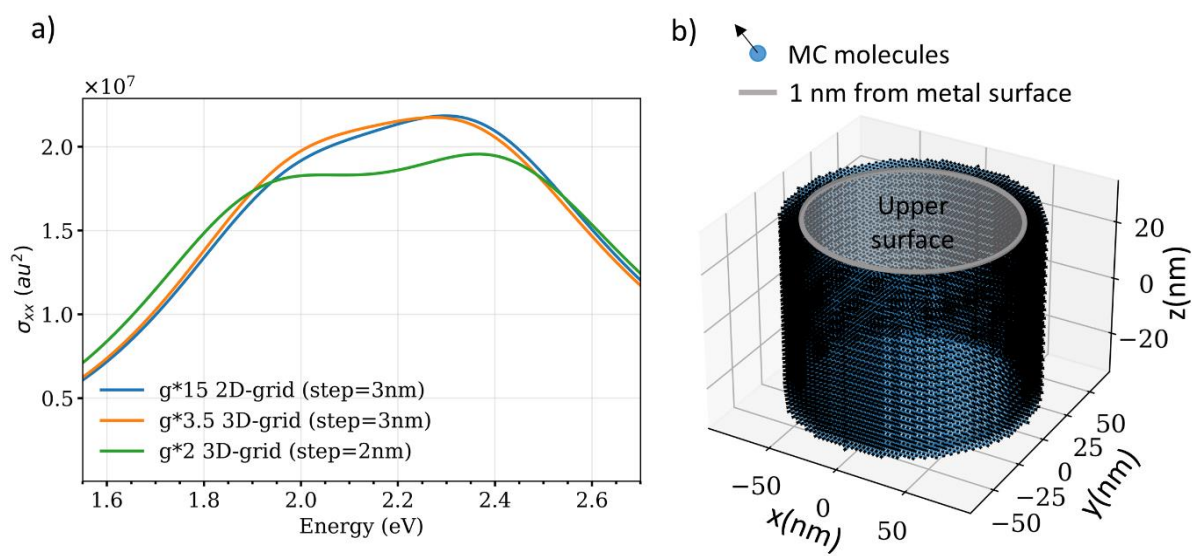

*Figure S7:* a) Simulated linear absorption of the polaritonic system (LA case) considering different grids and scaling factors for the computed couplings: 2D grid only (step size 3 nm between nearest neighbors and couplings scaled by a factor of 15) (blue line), 3D grid (step size 3nm and couplings scaled by a factor of 3.5) (orange line) and 3D grid (step size 2nm and couplings scaled by a factor of 2). b) Scheme of the 3D grid employed to obtain the orange line panel a.

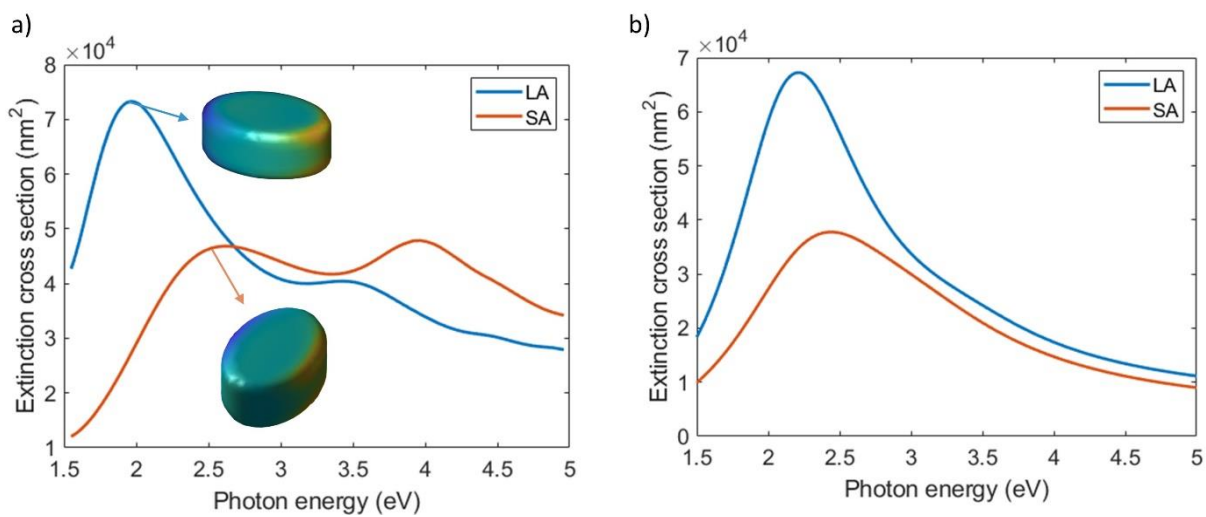

***Figure S8:*** *a) Simulated extinction spectra of the model nanoellipse computed by solving the full Maxwell equations through the MNPBEM code[14] for two distinct electric field polarizations of the incoming wave: long-axis direction (blue line) and short-axis direction (orange line). The aluminum dielectric function employed in the simulations is that of McPeak et al. [15] and the environment refractive index was set to 1.5. The two small insets display the surface charge distribution associated with the corresponding main peaks (dipolar plasmon resonances of interest); the yellow color represents positive charges, whereas blue represents negative ones. The two peak energies are very close to the experimental ones (2.15 and 2.45 eV after Spy deposition, Figure 1 panels c-d main text). b) Simulated quasistatic extinction spectra of the same nanoellipse by adopting a single Drude-Lorentz oscillator model for the dielectric function and tuning the corresponding parameters to reproduce the principal features of the two main dipolar resonances shown in panel a). The plasma frequency was set to 0.025 a.u. (eq. 5) and the following decay rates $\Gamma_{LA}$=0.038 a.u. and $\Gamma_{SA}$=0.053 a.u. for the long-axis (LA) and short-axis (SA) plasmons were employed.*